**Coherent coupling dynamics in a quantum dot microdisk laser**

D.K. Young, L. Zhang, D. D. Awschalom, and E. L. Hu

*Center for Spintronics and Quantum Computation*
*University of California, Santa Barbara, CA 93106, USA*

Abstract

Luminescence intensity autocorrelation (LIA) is employed to investigate coupling dynamics between (In,Ga)As QDs and a high-Q (~7000) resonator with ultrafast time resolution (150 fs), below and above the lasing threshold at T = 5 K. For QDs resonant and non-resonant with the cavity we observe both a six-fold enhancement and a 0.77 times reduction of the spontaneous emission rate, respectively. In addition, LIA spectroscopy reveals the onset of *coherent* coupling at the lasing threshold through qualitative changes in the dynamic behavior and a tripling of the resonant QD emission rate.

The ability to control spontaneous emission[1] in the solid state is expected to improve laser devices[2] as well as enable novel systems for quantum cryptography, such as single photon sources.[3] 'Self-assembled' InAs quantum dots (QD)s integrated within high-Q microcavities provide an ideal testbed for studying the control of QD emission through cavity QED (Purcell effect).[4,5] Moreover, understanding the coupling dynamics between the QDs and microcavity may provide an avenue for performing quantum computation and communication with electron spins.[6] Previous measurements in microcavities (i.e. micropillars,[4] microdisks,[5] and microspheres[7]) have explored enhanced spontaneous QD emission (relative to the unprocessed material), and have yet to investigate the coupling dynamics near the onset of lasing.

Here we report time-resolved measurements in microdisks with integrated QDs below and above the lasing threshold using luminescence intensity autocorrelation (LIA) spectroscopy,[8] where the time resolution is theoretically limited by the optical pump/probe pulse widths (~ 150 fs). Detailed studies of the QD-cavity coupling dynamics have shown for QDs resonant and non-resonant with the cavity, a six-fold enhancement and a 0.77 times reduction of the spontaneous emission rate, respectively. Moreover, we use LIA spectroscopy to measure the onset of *coherent coupling* between the QDs and cavity at the lasing threshold, where we observe a clear signature at zero time delay between pump and probe pulses and an additional tripling of the resonant QD emission rate due to stimulated emission.

The device structure is grown by molecular beam epitaxy upon a semi-insulating GaAs substrate with an AlAs/GaAs buffer. A 1 μm layer of $Al_xGa_{1-x}As$, (x ranging from 0.65-0.85) is subsequently grown, determining the microdisk post height. The disk region has a single layer of (In,Ga)As self-assembled QDs with a dot density of $10^9 cm^{-2}$ and is clad symmetrically by 100 nm GaAs, 20 nm $Al_{0.3}Ga_{0.7}As$, and 4 nm GaAs [right inset of Figure 1(a)]. The microdisks have ~ 4 μm diameters defined by photolithography, yielding an effective modal volume $V \sim 10(\lambda/n)^3$ [assuming the dominating modes are guided primarily around the circumference of the microdisk[9]; i.e. whispering gallery modes (WGM)s], where n is the effective refractive index of the cavity and $\lambda$ is the emission wavelength. WGMs arise due to total internal reflection, relying on the large index contrast between the semiconductor/air interface, where light is reflected along the curved inner boundary of the cavity.[10] Details of the two-step wet etch process and stimulated emission from QDs integrated in microdisks can be found elsewhere.[9] In order to minimize thermal broadening of the QD emission, all data is taken at T = 5 K in a liquid helium-flow optical cryostat and is collected from the side of the disk to maximize the collection efficiency of the WGM emission. Time-integrated photoluminescence (PL) is spectrally analyzed with a 0.5 m spectrometer coupled to a $LN_2$ cooled charge coupled detector yielding resolution of 50 μeV.

The PL data taken on unprocessed (i.e., as-grown) QD material [Figure 1(a) left inset], shows the characteristic broad spectrum at various pump intensities denoting the variation in size (13%) of the QDs. Increasing excitation intensity gives rise to increased emission at higher energies, suggesting occupation of, and subsequent emission from, higher energy QD transitions. Non-resonant excitation ($E_{ex}$ = 1.57 eV) into the GaAs layer is used to increase carrier generation due to the small absorption cross-section of the QDs.[9] Subsequent relaxation into the QDs occurs on very short time scales (carrier capture time ~ 20 ps).[5] Figure 1(b) shows multi-mode lasing under similar conditions of QDs within a microdisk (note semi-log scale). Indeed, PL from the microdisk is shown above lasing threshold on a linear scale [Figure 1(a)]. Quality factors, Q, as high as 7000 are determined by computing $Q = E/\Delta E$, where $\Delta E$ is the FWHM of the emission energy. The enhancement of the spontaneous emission rate is given by the Purcell factor,[1] $F_P$, (which indicates the degree of coupling between emitter and cavity): $F_P = (3/4\pi^2)(Q\lambda^3/V)$. For our highest Q WGM, with transition energy at 1.378 eV, we estimate $F_P \sim 50$. This calculation assumes an *ideal* monochromatic emitter that is spatially and spectrally coupled to the mode, however due to non-idealities in coupling between emitter and cavity as well as cavity mode degeneracy,[5] we expect the actual $F_p$ measured through the spontaneous emission lifetime to be smaller.

We employ luminescence intensity autocorrelation,[8] a time-resolved technique that measures the non-linear dependence of PL with excitation intensity to obtain the radiative lifetime of the QD excitonic states. LIA data are obtained using two energetically degenerate 150 fs pulses from a Ti-sapphire laser where relative time delay is controlled by a mechanical delay line. Optical choppers modulate the two equally intense excitation beams at different frequencies while phase-sensitive detection at the sum frequency measures non-linear changes in PL intensity that arise from the temporal overlap of the carrier populations excited by the two pulses. The normal incident excitation is focused to a spot (~30 μm) that is much larger than the microdisk diameter insuring uniform excitation of the cavity. PL in the time-resolved data is spectrally analyzed with a high throughput spectrometer with spectral resolution (~5 meV) and detected with an InGaAs photodiode.

Each point in the LIA scan is the *time-integrated* non-linear PL response for a given time delay between the two excitation beams. Moreover, there are two major contributions to the signal; (a) differential absorption and (b) “cross-recombination.” Differential absorption arises due to state filling, (i.e. the excitation from the first beam makes the second beam less likely to excite as many carriers, due to fewer available states), resulting in a negative LIA signal. In cross-recombination, electrons/holes from one pulse recombine with holes/electrons from the other, and result in a positive LIA signal. In our measurements, the PL intensity from one pulse increases in the presence of the second pulse, indicating that the LIA is determined by the radiative cross-recombination of the excitons and provides a measure of their lifetimes.

To compare emission rates, we measured the intrinsic lifetimes of QDs in an unprocessed part of the sample. Figure 2 shows the intensity dependence of LIA data taken at a detection energy $E_d$ = 1.322 eV, a ground state energy within the QD ensemble. LIA data are fit to a single-exponential $A \exp(-t/\tau)$ to extract the decay time of the recombining excitons. Since the beams are equal in intensity,[11] LIA data is expected to be symmetric around t = 0, as is observed. The data is well fit by a single-exponential, yielding lifetimes that agree with conventional time-resolved PL measurements.[12] For all extracted lifetimes, the error is smaller than the plotted symbol size or otherwise noted by error bars. QD spontaneous emission rates have been observed to decrease with increasing QD size[12] despite theoretical predictions that the QD oscillator strength should remain nearly independent of QD size.[13] We observe similar behavior in the LIA lifetimes of our unprocessed control samples (Figure 2 inset), whose insensitivity to pump intensity suggests they measure the QD spontaneous emission rate.

In contrast, the data from processed microdisk samples shows completely different behavior due to the presence of WGMs. The intensity dependence of the time-integrated PL from QDs within the microdisks is shown in Figure 1(c). Transitions coupled to the WGMs show non-linear dependence of PL intensity on excitation intensity, revealing lasing thresholds ($I_{TH}$ ~ 150 $Wcm^{-2}$ - 200 $Wcm^{-2}$). In contrast, for QDs off-resonant with a cavity mode (○; $E_d$ = 1.366 eV), there is a linear dependence on pump intensity [shown also in an expanded view in the inset]. In addition, at lasing threshold, the quality factor Q is largest and monotonically decreases with higher pump intensity [Figure 1(d)], while below threshold, Q ~1. Similar broadening in quantum well microdisk cavities at intensities above lasing threshold[14] is typically associated with free carrier absorption (FCA), however QD emitters in microdisk cavities with continuous wave excitation do not suffer from FCA at lasing threshold.[9] The decrease in Q may reflect the increased number of QDs becoming resonant with the cavity mode at higher intensities, as well as emission broadening from shorter lifetimes, or increased carrier densities with pulsed excitation.[15]

In Figure 3(a), normalized LIA data from QDs resonant (□) and non-resonant (○) with a WGM (Q ~ 7000), as well as energetically degenerate (ED) QDs (■) in an unprocessed part of the sample are compared for a given intensity (I = 140 $Wcm^{-2}$) near lasing threshold of E = 1.378 eV. PL from the microdisk under the same conditions is shown (inset). There is a large reduction in spontaneous emission lifetime from QDs that are resonant with the WGM (□; $\tau$ = 70 ps) when compared to both ED QDs outside the cavity (■; $\tau$ = 430 ps) and non-resonant QDs in the cavity (○; $\tau$ = 700 ps). The decay times of the non-resonant (○) QDs are relatively long in comparison to the resonant (□) QDs, indicating that the significant reduction in lifetime is not due to enhanced non-radiative recombination from sample processing. Indeed, a six-fold shortening of the lifetime as compared to ED QDs outside the cavity reflects the modified vacuum field of the microdisk (Purcell effect),[1] revealing non-idealities in the coupling (note $F_P$ ~ 50). A summary of lifetimes extracted from LIA data taken under non-lasing conditions is shown in Fig. 3(b). The *filled* symbols are from QDs in unprocessed regions, while *unfilled* symbols are from QDs in the microdisk. Note that spontaneous emission lifetimes from resonant QDs can only be unambiguously extracted below their lasing thresholds. This will be addressed in the Figure 4 discussion. Upon further review of Figure 3(b), the intensity dependence from QDs that are non-resonant with the cavity (○; $E_d$ = 1.366 eV) is quite different than ED QDs outside the microdisk (●). Near lasing threshold of a neighboring mode (E = 1.378 eV), non-resonant QDs *in the cavity* have longer lifetimes than ED QDs *outside the cavity*, indicating an inhibited spontaneous emission rate. However, above threshold, QD lifetimes *inside the cavity* begin to shorten. The transition at threshold may be due to the increased average number of photons in the cavity ($N_P$) when lasing. Above threshold, emission rates should be further enhanced due to the increased number of photons in the cavity following the relation $\Gamma_{TOT} \approx \beta_{sp}F_P\Gamma_{sp} + v_g g_{th} N_P$, assuming negligible non-radiative recombination where $\Gamma_{TOT}$ is the total emission rate, $\beta_{sp}$ is the spontaneous emission factor, $\Gamma_{sp}$ is the intrinsic spontaneous emission rate, $v_g$ is the group velocity of the mode of interest and $g_{th}$ [$cm^{-1}$] is the gain at threshold.[16]

We now focus on the coupling between resonant QDs and a high-Q WGM at the onset of lasing. Figure 4(a) shows intensity dependent LIA data for QDs detected at a WGM energy, $E_d = 1.378$ eV. Clearly, the large dip at t = 0 delay in the LIA data is markedly different from the intensity dependence of QDs in the unprocessed material [Fig. 2]. The 0 delay dip manifests in a shift of the peak position in the LIA signal from 0 to 86 ps, predominantly occurring at the lasing threshold. This non-monotonic form of the LIA signal is not observed at similar pump intensities (> 400 $Wcm^{-2}$) in non-resonant QDs within the microdisk (○) *or* in ED QDs in unprocessed material (■) [Figure 4(b)], suggesting that the dip at 0 delay is associated with stimulated emission, a signature of *coherent* coupling between QDs and WGM. Moreover, WGMs at $E_d = 1.387$ and 1.396 eV in this microdisk as well as other WGMs in other microdisks, show similar behavior at 0 delay above their lasing thresholds (not shown). The dip in the LIA signal may be due to a suppression of cross-recombination due to the enhanced emission rate under stimulated emission.[16] However, further investigation into the mechanism is needed. We now fit the right-most edge of the LIA signal to a single-exponential. Note that the fit is initiated at later times as the pump intensity is increased [shown in Figure 4(a)]. Indeed, we observe a further enhancement (3×) in the resonant QD emission rate due to the increase in average photon density, $N_P$, by stimulated emission. The dip 'half-width' at 0 delay as well as the saturation of the extracted lifetime ($\tau$ ~ 25 ps) at higher pump intensities may be limited by one or the combination of processes with comparable timescales: carrier capture time (~20 ps),[5] photon lifetime in the cavity ($Q/\omega$ ~ 3 ps), or stimulated emission lifetime.

Finally, we estimate the number of QDs coupling to the cavity. Given the QD density and diameter of the microdisks, we can estimate the total number of QDs in a microdisk (~200). Considering the Q of the cavity mode at lasing threshold (~5000) and neglecting the QD's lifetime dependence on emission energy, we can estimate the average number of QDs contributing to the WGM from the PL of the unprocessed region to be ~2. However, at pump intensities needed for lasing, homogeneous broadening (~5 meV) of a QD ground state transition[17] may allow on average ~25 QDs to weakly couple to a WGM due to partial spectral or spatial overlap. Using this value, the average photon number in this mode[5] is $N_P \sim 1.9$, indicating lasing action. Additionally, we find $\beta_{sp} < 1$ (~ 0.43), consistent with an observed lasing threshold. In summary, LIA spectroscopy is employed to study the *coherent* coupling between QDs and a high Q WGM in a single microdisk at the onset of lasing, revealing rich dynamics in the QD-cavity coupling and showing promise for studying ultrafast phenomena in novel solid-state systems for quantum information processing.

The authors would like to thank R. J. Epstein, J. A. Gupta, A. Imamoglu, and J. Levy for inspiring discussions. Work supported by the AFOSR F49620-02-1-0038, NSF DMR-0071888, and DARPA/ONR N00014-99-1-1096.

**Figure Captions**

Figure 1

Time-integrated photoluminescence (PL) characteristics. (a) Multi-mode lasing from a high-Q microdisk integrated with QDs. (Left inset) Intensity dependent QD PL from unprocessed material, (Right inset) Band diagram showing excitation into GaAs and subsequent relaxation into QD ground and excited states. (b) Intensity dependent PL from the microdisk. (c) Non-linear PL intensity dependence for three whispering gallery modes (WGM)s. Symbols defined in (b). QDs non-resonant with the cavity show linear dependence. (Inset) Expanded view. (d) WGM quality factors Q measured at intensities above their lasing thresholds.

Figure 2

Time-resolved QD PL characteristics in unprocessed material. Intensity dependent luminescence intensity autocorrelation (LIA) data detected at QD energy ($E_d$ = 1.322 eV). (Inset) Lifetimes extracted from decaying single-exponential fits (solid lines) are plotted as a function of pump intensity (●) for fixed $E_d$, and $E_d$ (◊) for fixed pump intensity (300 Wcm-2).

Figure 3

(a) A six-fold enhancement in the spontaneous emission rate is observed when comparing normalized LIA data between QDs resonant with a WGM (□) to energetically degenerate (ED) QDs in unprocessed material (■). QDs non-resonant with a WGM (○) under same conditions indicate enhancement is indeed due to Purcell effect. (Inset) PL taken under same conditions. Shaded regions represent spectral resolution in time-resolved measurements. (b) A summary of below threshold lifetimes for QDs outside the cavity (filled symbols), as well as QDs resonant and non-resonant with a WGM (unfilled symbols). Below lasing threshold of a nearby mode, inhibited spontaneous emission is observed when comparing QDs non-resonant (○) within the cavity to the ED QDs (●) outside the cavity.

Figure 4

(a) Intensity dependent LIA data for QDs resonantly coupled to a WGM driven above lasing threshold ($I_{TH}$ ~140 $Wcm^{-2}$). (Inset) Same LIA data plotted from –250 ps to 250 ps. (b) Normalized LIA data detected at QD emission non-resonant with the cavity, as well as ED QDs in the unprocessed region taken under high intensity (570 and 400 $Wcm^{-2}$, respectively), shows no dip at 0 delay. (c) Fitted lifetimes from right most edge of data in (a) indicate a tripling of the resonant QD emission rate due to stimulated emission.

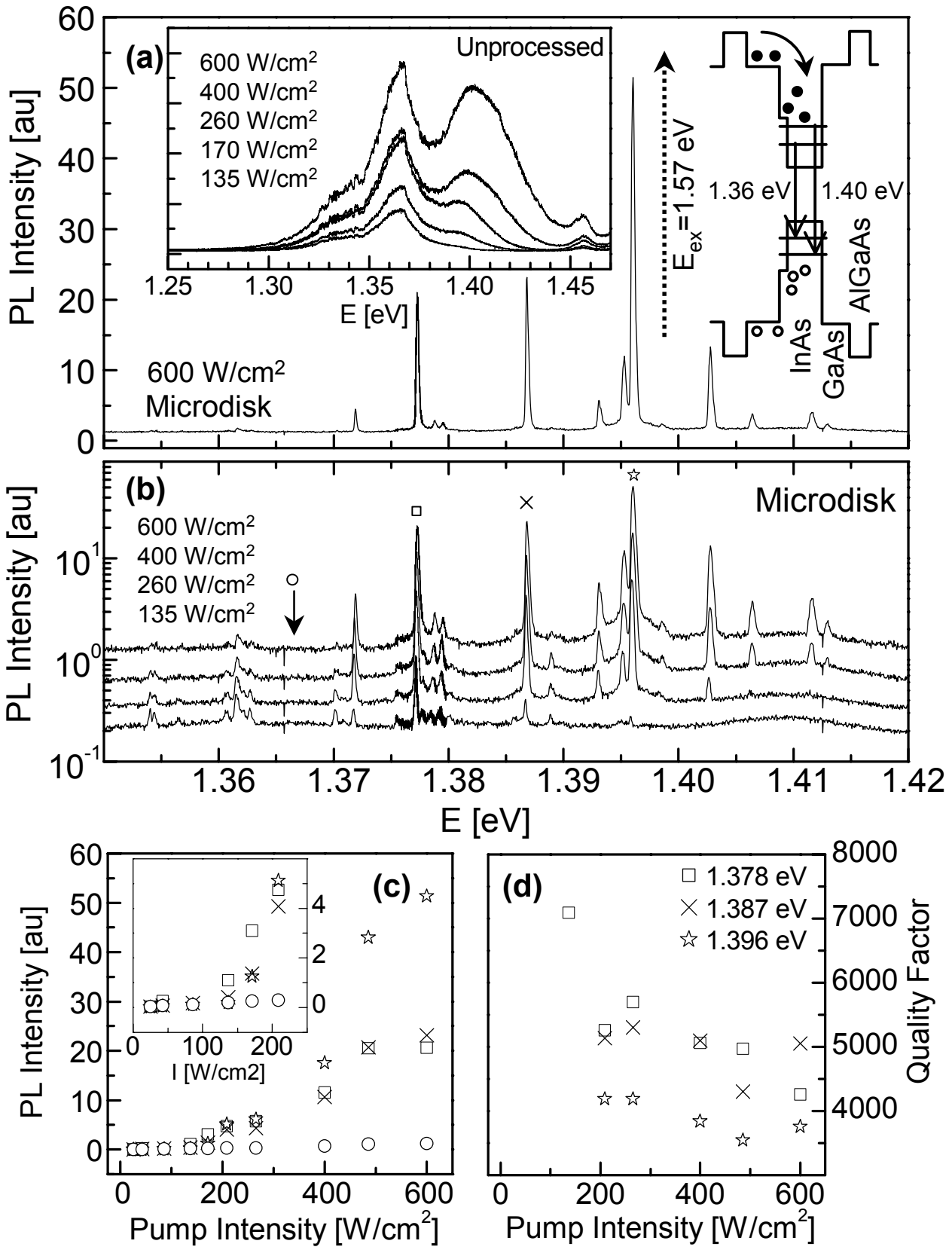

Figure 1 Young et al

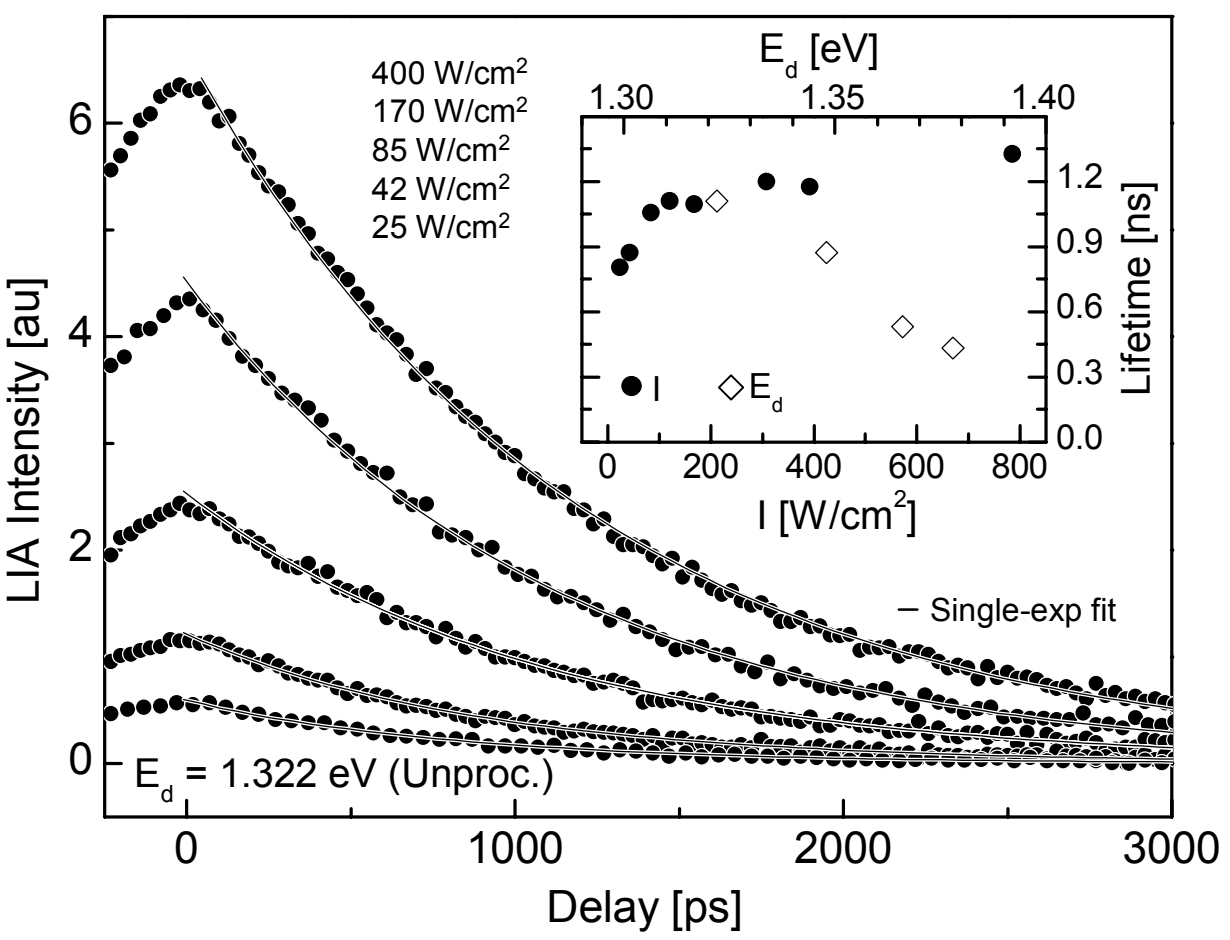

Figure 2 Young et al

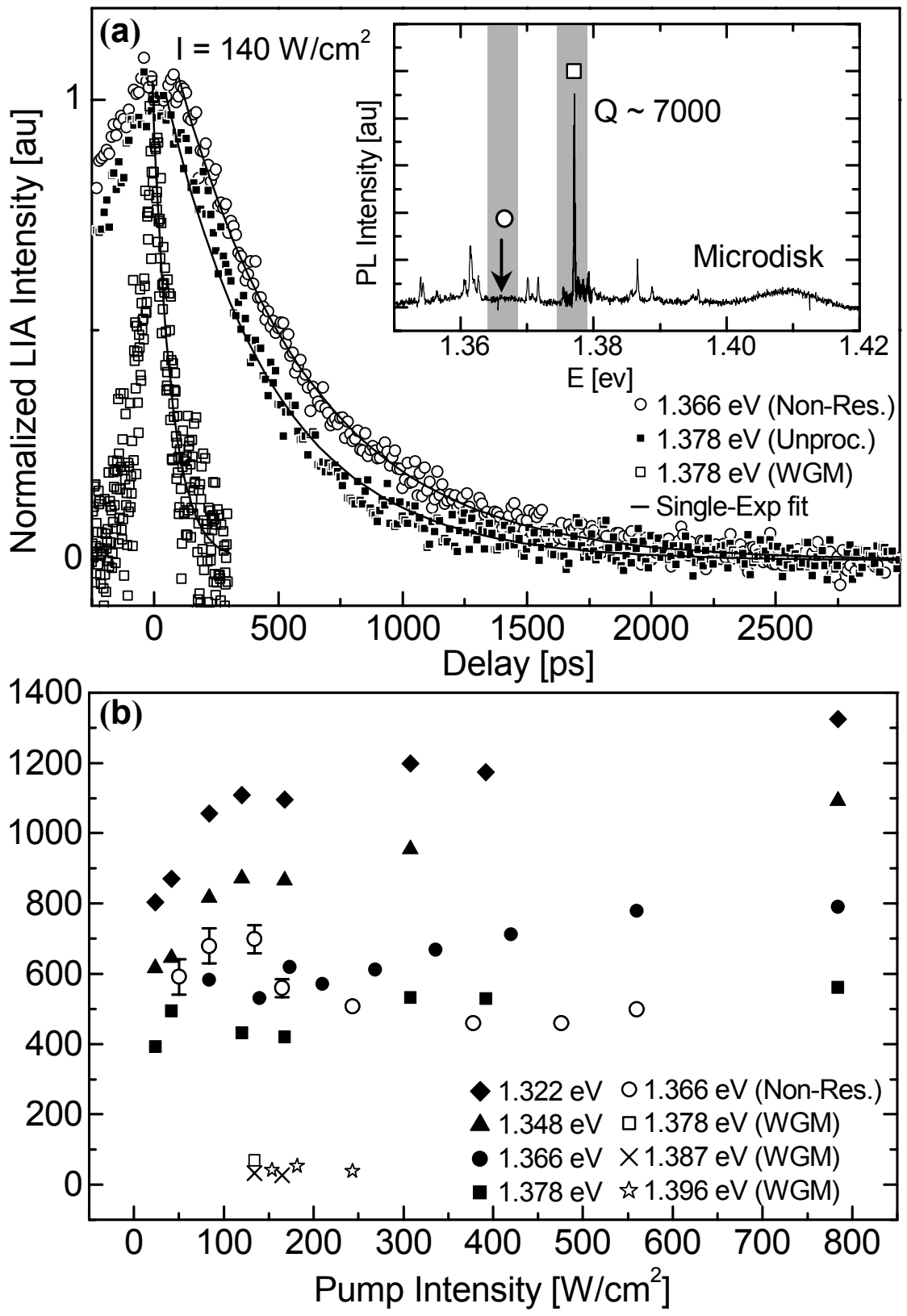

Figure 3 Young et al

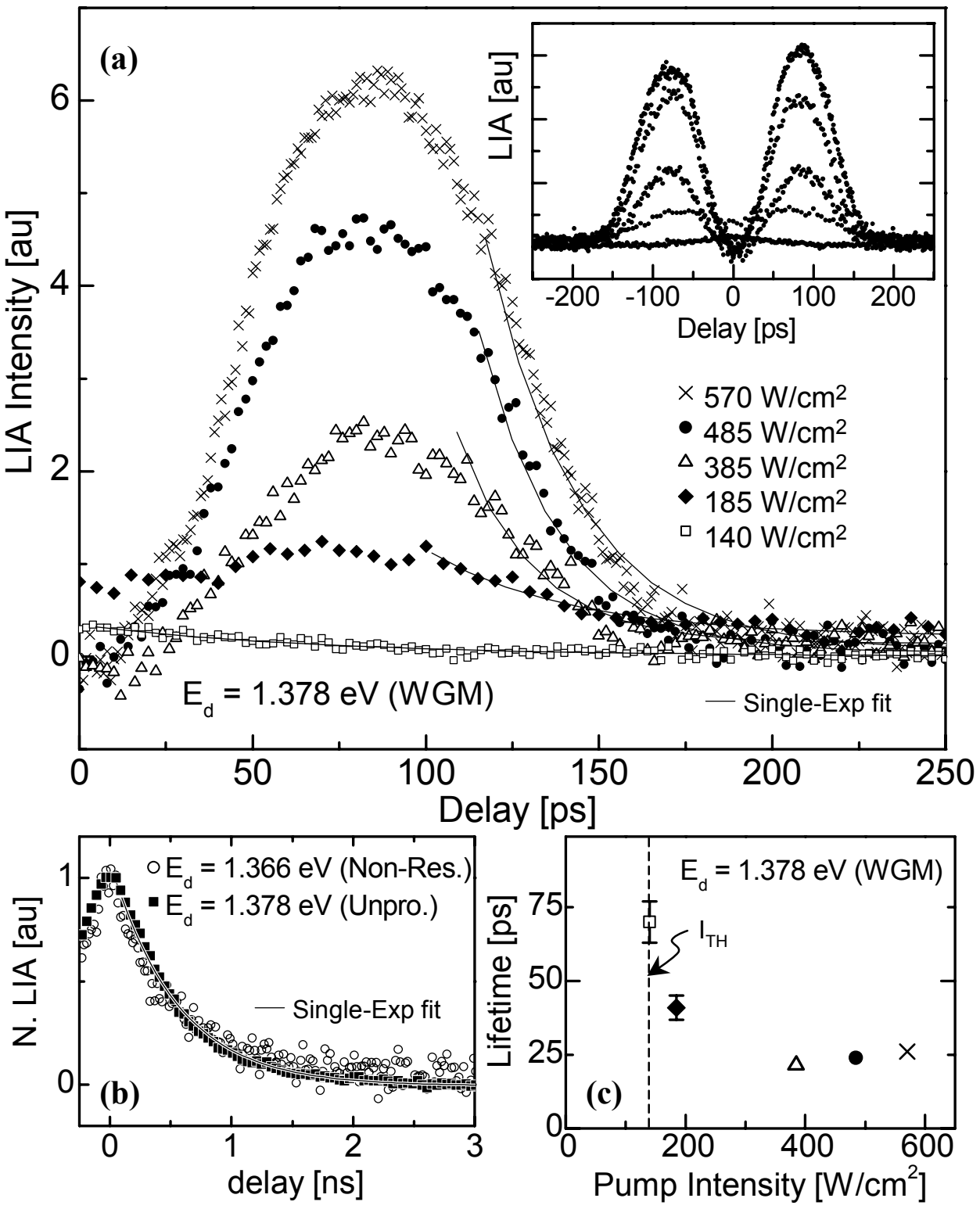

Figure 4 Young et al